\newcommand{\vy}{{\vec{y}}}
\newcommand{\vb}{{\bf b}}
\newcommand{\vs}{{\vec{s}}}
\newcommand{\vomega}{{\vec{\omega}}}

\newcommand{\rmd}{{\bf d}}

\newcommand{\cut}[1]{}
\newcommand{\PREPRINTYES}[1]{#1}
\newcommand{\PREPRINTNO}[1]{}
\newcommand{\UPDATEBIBFILEYES}[1]{}
\newcommand{\UPDATEBIBFILENO}[1]{#1}
\documentclass[conference]{IEEEtran}
\ifCLASSINFOpdf
   \usepackage[pdftex]{graphicx}
\else
   \usepackage[dvips]{graphicx}
\fi
\usepackage[cmex10]{amsmath}
\usepackage{amssymb}
\begin{document}
%
\title{Optimal sparse CDMA detection at high load}

\author{\IEEEauthorblockN{Jack Raymond}
\IEEEauthorblockA{School of Physics\\
Hong Kong University of Science and Technology\\
Hong Kong,\\
Email: jack.raymond@physics.org}
}



%


\maketitle

\begin{abstract}
Balancing efficiency of bandwidth use and complexity of detection involves choosing a suitable load for a multi-access channel. In the case of synchronous CDMA, with random codes, it is possible to demonstrate the existence of a threshold in the load beyond which there is an apparent jump in computational complexity. At small load unit clause propagation can determine a jointly optimal detection of sources on a noiseless channel, but fails at high load. Analysis provides insight into the difference between the standard dense random codes and sparse codes, and the limitations of optimal detection in the sparse case.
\end{abstract}


%
\IEEEpeerreviewmaketitle

\section{Introduction}

Multiuser detection is the problem of extracting estimations of multiple sources from a shared communication channel~\cite{Verdu:MD}. In order to most efficiently use the bandwidth a high load should be used, and many theoretical results indicate achievable capacities are good in this regime for standard code classes. However, increasing load forces users to share bandwidth and creates a strongly correlated inference problem, for which optimal detection may not be possible by practical (fast) detectors. Even in the limit of zero noise in the channel, multi-access interference may prevent a Jointly or Individually Optimal (JO/IO) estimation of the sources.

Code Division Multiple Access (CDMA) is a method of bandwidth allocation in which each of $K$ users is assigned a code ($\vs_k$) by which to modulate a symbol on the bandwidth of $\beta K$ ($M$) orthogonal time/frequency blocks (chips), $\beta$ is called the load.
The scenario of a noiseless multi-access channel is examined in this paper for ensembles of sparse random codes. The sparse codes examined have the advantage that they can be assigned independently at random to all users, and are known to have a good performance in Additive White Gaussian Noise Channels (AWGNC), combined with Belief Propagation (BP) decoding~\cite{Yoshida:ASS,Montanari:BPB,Raymond:SS}.

Using only Unit Clause Propagation (UCP)~\cite{Deroulers:CUU} all source may be determined efficiently for $\beta$ up to some discontinuous transition point beyond which detection by decimation becomes suddenly inefficient. 

The noiseless threshold results provide bounds on achievable detection and may guide the development of algorithms. Features of this transition may be relevant to decimation based detectors in a variety of sparsely coded noisy channels. For AWGNC at standard operating power levels (signal to noise ratio $\sim \! 6-9$dB) the departure of the signal from the noiseless case is quite small and threshold behaviour in the noiseless system may have a dominating effect on detector performance.

\subsection{CDMA model and pseudo-random codes}
\begin{figure}[!htp]
\centering{
\includegraphics[angle=270,width=\PREPRINTNO{0.85\linewidth}\PREPRINTYES{\linewidth}]{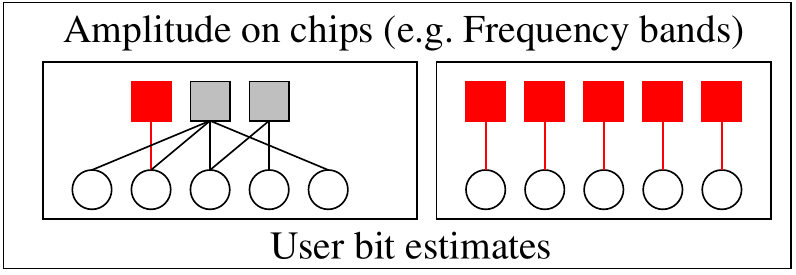}
\caption{\label{fig:figure1} (colour online) Two different coding schemes are demonstrated with $K\!=\!5$ and load $\beta\!=\!1$ (right) and $5/3$ (left). Connections indicate a non-zero transmission by the user on the chip ($s_{\mu k}\!=\!\pm 1$). Left: When several users are coincident on one chip inference of one variable depends on values of the others. Right: Since each user has a unique chip detection is much easier, but this scenario requires coordination of user transmissions and small load $\beta$.}
}
\end{figure}

A standard synchronous CDMA model has each user transmitting a modulated bit $b_k\!=\!\pm 1$, the sources interfere to give a signal
\begin{equation}
\vy=\sum_k \vs_k b_k +\vomega\;, \label{eq:vy}
\end{equation}
where $\vomega$ represents channel noise. Different encodings may be represented as graphical inference structures, as shown in figure \ref{fig:figure1}. Weakly correlated codes, such as orthogonal codes create computationally easy detection problems, at high loads, or with poorly chosen codes, detection may be hard.

For many noise models, including the noiseless limit, a maximum capacity can be achieved by minimising overlaps in user codes $\vs_k.\vs_l$~\cite{Rupf:OSM}, orthogonal codes are a solution when $\beta \!\leq\! 1$. However, creating maximum distance codes is computationally expensive as $\beta$ increases, and the allocation of codes can be inflexible.

In realistic operating conditions synchronisation of users may not be possible, and codes must be made robust against a number of phenomena. Furthermore a small loss in capacity might be tolerable in order to achieve greater flexibility in the code allocation, or efficiency in the detection process. For this reason codes sampled independently from code ensemble are often considered. Random codes in which every user accesses $M$ chips with a unique modulation pattern imply inference structures described by dense graphs~\cite{Tanaka:SMA}, and these have become favoured in theory and practice. More recently it has been argued that sparse codes, where each user accesses only $C$ ($\ll M$) chips might have some favourable properties~\cite{Yoshida:ASS}, particularly due to the efficiency of BP and message passing.

Analysis by Tanaka~\cite{Tanaka:SMA} suggests that for dense codes in the asymptotic case of many user and large bandwidth, limits of $\beta_c\!=\!2.09 (1.51)$ exist above which IO (JO) detection is not expected to be efficient due to suboptimal attractors for detector dynamics, even in the noiseless limit. In a variety of experiments based on BP, and heuristic decoding, the $\beta\gtrsim 1$ regime indeed proves to be difficult~\cite{Montanari:BPB,Raymond:Thesis} for sparse codes. The majority of practical detectors have efficient working regimes restricted to $\beta \lesssim 1$.

In this paper the origins of this hardness at large load are investigated for sparse codes in a noiseless channel. UCP is found to be sufficient to produce a jointly optimal detection of the bits in some range of $\beta$, so that perfect detection is possible for loads up to some critical threshold depending on user connectivity. The breakdown of the UCP detector is sudden and leaves a residual problem without an obvious solution, the residual problem has many properties characteristic of hard constraint satisfaction problems~\cite{Zdeborova:LC}.

\subsection{Sparse random codes}
The set of codes examined are sparse so that each user transmits on only a fraction $C/K$ of chips, Binary Phase Shift Keying (BPSK) is used so the marginal is
\begin{equation}
P(s_{\mu k}) \!=\! \left(1- \frac{C}{K}\right)\delta(s_{\mu k}) + \frac{C}{K}\frac{\left(\delta(s_{\mu k}-1)+\delta(s_{\mu k}+1) \right)}{2} \label{eq:Poissonian} \;,
\end{equation}
each non-zero transmission is a binary modulation $\pm 1$. The degeneracy problem central to this paper can be avoided by choosing another modulation pattern, but degeneracy becomes a problem for such schemes as soon as realistic noise is introduced.

In the Poissonian ensemble each chip is accessed in an independent manner, unfortunately some users end up transmitting on no chips in this ensemble. In the more practical Regular ensemble all users access exactly $C$ chips so that
\begin{equation}
P(\vs_k) \propto \delta(\sum_{\mu=1}^M (s_{\mu k})^2- C) \prod_{\mu=1}^M  P(s_{\mu k}) \label{eq:regular}\;.
\end{equation}
In the limit of large $M$ the case $C\!=\!O(1)$ has properties distinguishable from the dense case due to dilution effects.

\section{Detection by Unit Clause Propagation}
\cut{\begin{figure}
\includegraphics[width=\PREPRINTNO{0.85\linewidth}\PREPRINTYES{\linewidth}]{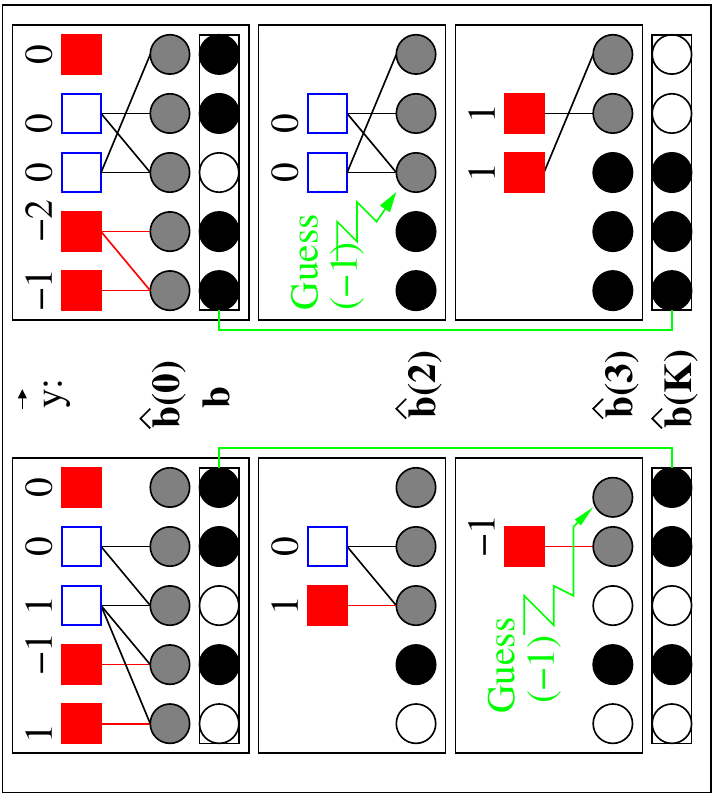}
\caption{\label{fig:figure2} (colour online) Application of the algorithm is demonstrated for two small noiseless channels. A set of bits black (-1) and white (+1) are modulated onto a channel producing a signal $\vy$. Estimates to the bits, initially neutral (grey) are formed by unit clause propagation. The signal includes some non-degenerate clauses (3 in both cases), implying 2 variables. From this initial condition variables are decimated from the set of unit clauses creating new implications. In the left figure clauses which are initially degenerate become non-degenerate with the inclusion of new information. In both graphs some variables cannot be inferred and must be guessed, in the right sequence guessing of one variable implies the value of another. Finally an estimate after $K$ decimations is found, which is to be compared with the encoded bit sequence to determine a bit error rate. Since both sequences complete without a contradiction the solutions are Maximum-A-Posteriori solutions, although due to degeneracy in the inference problem the right solution is correct in only 2 out of 5 bits (the performance depends on the guess).}
\end{figure}
}
The inference problem in the noiseless channel consists of examining the signal and determining an estimation of the sent bits (${\hat \vb}$) consistent with the model (\ref{eq:vy}). The value of some bits might only be determined by looking at the entire signal, but other bits may be implied from only one chip. Each chip determined by $L$ user contributions, with signal $y$, can be interpreted as a logical clause. Some clauses allow degenerate solutions when considered in isolation, others imply unique solutions. If a user $k$ is the only transmission on chip $\mu$ for example, then $y_\mu\!=\!\pm 1$ and any consistent estimate requires ${\hat b}_k\!=\!s_{\mu k}y_\mu$, regardless of other chips on which $k$ transmits. The logical implication is a unit/atomic clause in the variable $k$. Similarly a chip combining $L$ interfering transmissions determines all the incident users (through $L$ unit clauses) if the signal is extremal $y\!=\!\pm L$. Otherwise the solution is degenerate, allowing several consistent assignments of incidents bits.

UCP produces an estimate by decimating variables contained in unit clauses and is a central process in many complete detectors. By iteratively removing the unit clauses the remaining degenerate clauses, which form a simplified inference problem (residual graph), may be modified and create additional unit clauses. If insufficient unit clauses exist at some point in the algorithm, one can guess a value at random, or using some more advanced inference, and UCP checks the logical implications of this guess. A consequence of guessing is that one may produce contradictory unit clauses at some later point in the algorithm, a consistent (JO) solution is not possible where contradictions occur. However, guessing may not need to coincide exactly with the bit sequence for a JO detection, it may be that several solutions exist for a given signal. This is an undesirable scenario for practical purposes, but possible at high load.

The initial set of unit clauses is found by taking all chips for which $y_\mu\!=\!\pm \sum_k |s_{\mu k}|$ and converting each to $|y_\mu|$ unit clauses. The set of unit clauses ${\bf \Omega}_+$ is populated, while this set is non-empty, and includes no contradictions, the algorithm proceeds deterministically in two steps.

In the first step a variable is decimated: Select a variable $k$ represented in ${\bf \Omega}_+$, set its value ${\hat b}_k$ according to the unit clause(s) and remove unit clauses in variable $k$ from ${\bf \Omega}_+$. Reevaluate the signal on all chips on which user $k$ transmitted: $y_\mu\rightarrow y_\mu-s_{\mu k}{\hat b}_k$, and then remove (set to zero) $s_{\mu k}$ for all $\mu$. Variable $k$ is removed from the problem leaving a residual graph. Let $X$ be the the number of variables assigned in this way, the decimation time, so that $X\rightarrow X+1$ in this step.

In the second step modifications to the residual graph are considered, chips formerly degenerate may now be informative. For every chip modified in the first step check whether $\sum_k |s_{\mu k}|\!=\!\pm y_\mu$, if this condition is met then the inference is no longer ambiguous and a unit clause is created and added to the set ${\bf \Omega}_+$, for each non-zero $s_{\mu k}$.

If ${\bf \Omega}_+$ is empty then a variable is chosen and a unit clause is created and added to the set ${\bf \Omega}_+$. In the simplest scenario the variable ($k$) and assignment ${\hat b}_k\!=\!\pm 1$ are chosen uniformly at random. Once the first such guess is made it is necessary to keep track of contradictions. If a unit clause is added to the set ${\bf \Omega}_+$ that is in contradiction with an existing unit clause the algorithm may proceed to termination by ignoring the new unit clause, but the estimate obtained is not JO.

The algorithm completes after $K$ decimations, either with the unique JO estimation (if no guesses were required), a JO estimation (if guesses were required, but no contradictions encountered), or an approximation (if contradictions were encountered). If a JO solution is required the algorithm may be modified so as to back-track and reevaluate previous guesses when contradictions are encountered.
\begin{figure}
\centering{
\includegraphics[width=\linewidth]{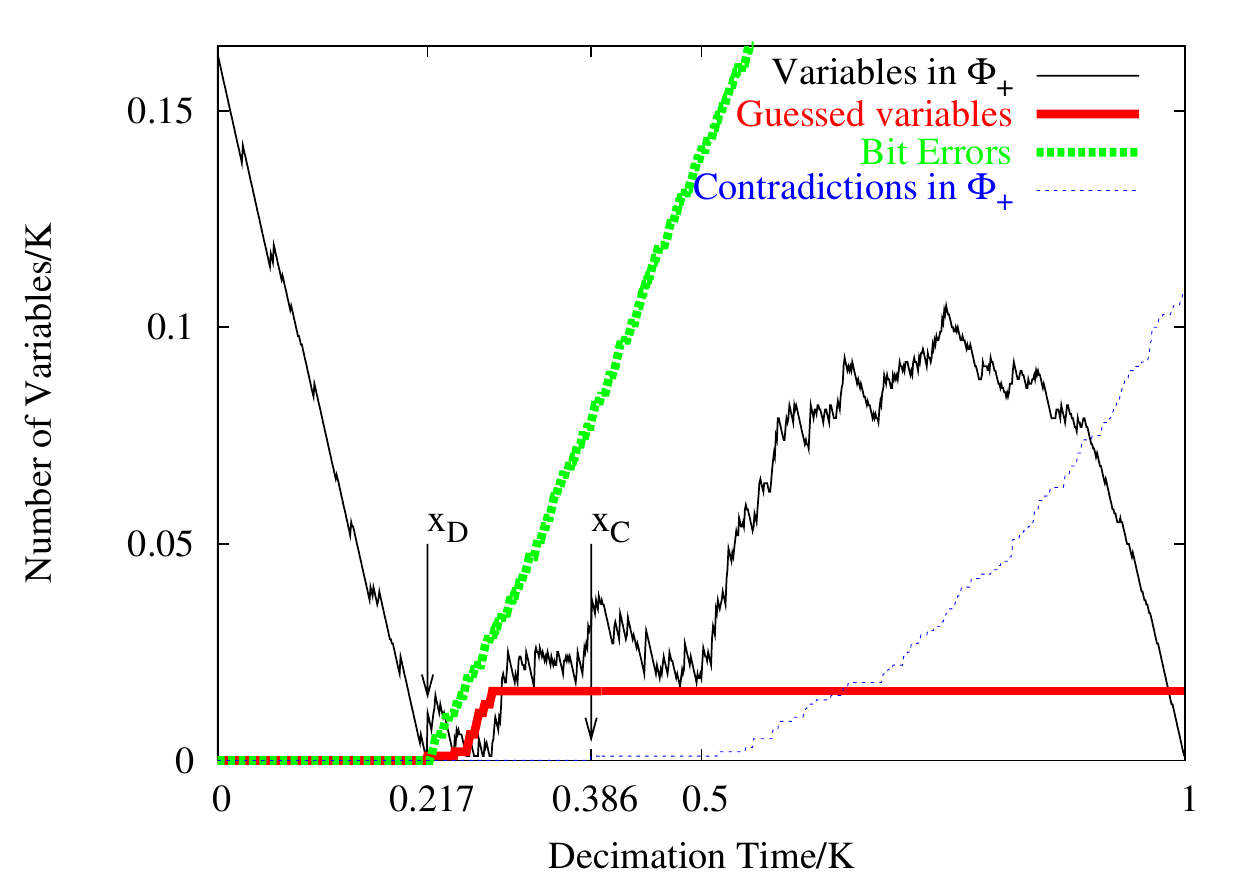}
\caption{\label{fig:Example63} (colour online) A typical decoding by the proposed method of a regular code transmission ($C\!=\!3$, $\beta\!=\!2$, $K\!=\!1000$) is shown. Up to decimation time $x_D K$ the process is deterministic, $|{\bf \Phi}_+|>0$, and errors are absent. Beyond this time a fraction of variables are determined by guesses, the bit error rate begins to increase linearly once the first guess is made. At decimation time $x_C K$ contradictions appear in ${\bf \Phi}_+$, indicating incorrect guesses. Both critical times exist in this example, it is also possible that the algorithm runs without requiring any guesses, or without producing contradictions, as occurs for smaller load.}
}
\end{figure}

An example of the algorithm in action for the third scenario is shown in figure \ref{fig:Example63} for an experiment with $C\!=\!3$, $\beta\!=\!2$, a random bit sequence, and codes sampled from the regular ensemble (\ref{eq:regular}). Up to a time $x_D K$ the algorithm is deterministic, and up to a time $x_C K$ no contradictions are encountered. Between $x_D K$ and $x_C K$ a small fraction of variables must be guessed, the Bit Error Rate (BER) increases linearly in expectation from the first guess. Many contradictions arise later in the algorithm, implied by the small number of guesses.

\section{Asymptotic Results}
\label{sec:asymptotic}
\begin{figure}[!htb]
\centering{
\includegraphics[width=\PREPRINTNO{0.85\linewidth}\PREPRINTYES{\linewidth}]{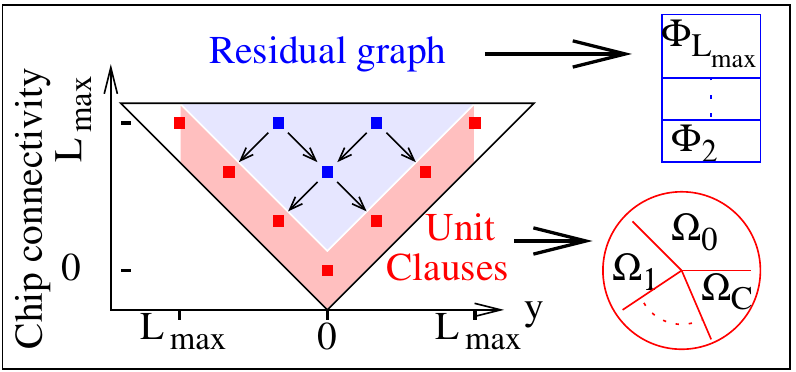}
\caption{\label{fig:graphstructure} (colour online) The initial inference problem consists of variables distributed at random within L-y clauses. In order to describe the mean properties of the algorithm at the initial conditions, and through the deterministic phase up to $x_D K$ decimations, $\Phi_l(X)$ and $\Omega_c(X)$ are sufficient statistics. These are respectively the number of clauses of length $l$ (ignoring clauses larger than $L_{max}$, a sufficiently large upper bound) describing the residual graph, and the number of variables appearing $c$ times in the set of unit clauses (${\bf \Omega}_+$), a sufficient description of ${\bf \Omega}_+$.}
}
\end{figure}

Experimentation indicates a self-averaging effect for large $K$ that can be analysed throughout the deterministic stage of the algorithm ($X/K<x_D$). This is possible by assuming a concentration process in the various extensive statistics of the residual graph~\cite{Deroulers:CUU,Raymond:PD}, and the extensive population of distinct unit clauses, the important extensive properties are illustrated in figure \ref{fig:graphstructure}. The Poissonian ensemble is analysed for brevity, the Regular ensemble derivation is more subtle and outlined in\PREPRINTYES{ Appendix \ref{sec:regularensemble}\footnote{This appendix is included here but absent in the Physcomnet submission for reasons of page limitation.}.}\PREPRINTNO{~\cite{Raymond:ODpreprint}.}

In the Poissonian ensemble the number of chips with $l$ incident users is Poissonian distributed, parameterised by $L$ ($\!=\!C\beta$), a finite upper bound is assumed for the largest relevant chip $L_{max}$. Of chips containing $l$ contributions, a fraction $2^{1-l}$ are non-degenerate for $l \geq 1$. The remaining clauses are degenerate, the number of degenerate clauses of length $l$ is called $\Phi_l(X)$. The fraction of signals taking value $y$ for given $l$ in the residual graph is Binomial in expectation at time $X\!=\!0$, and at later decimation times is found to be dependent only on $l$ given $\Phi_l(X)$ (a proof by iteration is possible). The initial condition is
\begin{equation}
\Phi_l(0)= \frac{K}{\beta}\left[\frac{\exp(-L)L^l}{l!} \right]\left(1 - 2^{1-l}\right) \;.\label{eq:Phil0}
\end{equation}
For the Poisson dynamics the only other statistic required is the number of un-decimated variables that are not represented in the set of unit clauses, $\Omega_0(X)$. For the Poissonian ensemble this number, at $X\!=\!0$, is the probability of a zero in the Poisson distribution parameterised by the total number of variables incident on chips of type $y\!=\!\pm L$.
\begin{equation}
\Omega_0(0) = K \exp \left\lbrace -\frac{1}{\beta} \sum_{l=1}^{L_{max}} l 2^{1-l} \left[\frac{\exp(-L)L^l}{l!} \right]\right\rbrace\;.\label{eq:Omega00}
\end{equation}
The number of times a variable is represented in the set of unit clauses, and the number of times it appears in the graph are conditionally independent at time $X\!=\!0$ given the extensive statistics. Therefore a variable decimated from the set of unit clauses is incident in chips of length $l$ a number of times proportional to length $l$, and inversely proportional to the number of remaining variables $K-X$. This clause may become either a smaller clause, or a set of $L-1$ unit clauses, the later with probability $z_l$ (independent of $X$). The population $\Phi_l(X)$ is modified according to
\begin{equation}
\begin{array}{lcl}
\Phi_l(X+1) &=& \Phi_l(X) - \frac{l}{K-X}\Phi_l(X) \\
&+& \frac{(l+1)}{K-X} (1-z_{l+1})\Phi_{l+1}(X) \;,\label{eq:DeltaPhi}
\end{array}
\end{equation}
except for $l\!=\!L_{max}$ where the second term is absent.
This applies intuitively at the first decimation time $X\!=\!0$, but it is also true at all $X$.
Variables which become unit clauses through the process of decimation do so at a rate proportional to their connectivity, and are represented in the residual graph with a frequency determined by their excess connectivity distribution. For the Poisson ensemble these are both the same and whether a variable is represented zero or several times in the set of unit clauses does not effect its distribution in the residual graph.

As the clauses are reduced a fraction become new unit clauses, determined by $z_l$ and $\Phi_l$. At time $X$, there are $K-X$ variables remaining, of which $\Omega_0(X)$ are unrepresented in a unit clause. Since the new unit clauses represent a random sample of the variables, $\Omega_0(X)$ has an expected change proportional to the fraction of variables it contains and the total number of new unit clauses
\begin{equation}
\Omega_0(X+1) = \Omega_0(X)\!-\! \frac{\Omega_0(X)}{K-X} \left[ \sum_{l=2}^{L_{max}} \frac{l(l-1)}{K-X} z_l \Phi_{l}(X) \right]\;.
\end{equation}

Writing the various terms at $O(K)$, ($X\!=\!xK$,$\Omega_0\!=\!\omega_0 K$,$\Phi_l\!=\!\phi_l K$), and Taylor expanding the left hand sides up to first order in $\frac{1}{K}$ leads to differential equations
\begin{equation}
\frac{\rmd \phi_l(x)}{\rmd x}= - \frac{1}{1-x}\phi_l(x) + \frac{(l+1)}{1-x} (1-z_{l+1})\phi_{l+1}(x) \label{eq:dphidx}\;,
\end{equation}
for $2\leq l <L_{max}$, for $l\!=\!L_{max}$ the second term is absent, and
\begin{equation}
 \frac{\rmd \omega_0(x)}{\rmd x}=  -\frac{\omega_0(x)}{1-x} \left[ \sum_{l=2}^{L_{max}} \frac{l(l-1)}{1-x} z_l\Phi_{l}(x)\right] \label{eq:domegadx}\;.
\end{equation}

These equations can be solved to determine the extent of the critical domain. It is necessary to find the first point in the interval $[0,1]$ where $\omega_0(x_D)\!=\!1-x_D$, this is the point where the set of unit clauses becomes empty. For the Poissonian code the equation (\ref{eq:dphidx}) is solved by a set of polynomials of the form
\begin{equation}
\phi_l(x) = \sum_{i=l}^{L_{max}} {\bf \phi}_{l,i} (1-x)^i x^{L_{max}-i}\;,
\end{equation}
where the coefficients can be determined iteratively from the equation for $l\!=\!L_{max}$, incorporated additional boundary conditions for smaller $l$. Using these solutions (\ref{eq:domegadx}) may be solved as an exponential form. Roots of $\omega_0(x_D)\!=\!1-x_D$ in the interval $[0,1)$ can be determined numerically.

Unfortunately the equation for the regular ensemble are not so easily solved. One must keep track of the dynamics for variables of different degeneracies in ${\bf \Omega}_+$, the number of variables appearing $c$ times as unit clauses is $\Omega_c(X)$\PREPRINTYES{, where $c$ runs from $0$ to $C$. A variable present in ${\bf \Omega}_+$ occurs less frequently in the residual graph, which must be accounted for in (\ref{eq:DeltaPhi}), and the new unit clauses contribute to the populations $\{\Omega_c(X)\}$ in a non-linear fashion. Including initial conditions one can again write a set of differential equations, but these must be solved by numerical integration. Fortunately the dynamics are smooth and fourth order Runge-Kutta methods seem to produce a good determination of the critical point, $x_D$.}\PREPRINTNO{. A comparable set of equations is established, but is solved only by numerical integration.}

\begin{figure}
\centering{
\includegraphics[width=\linewidth]{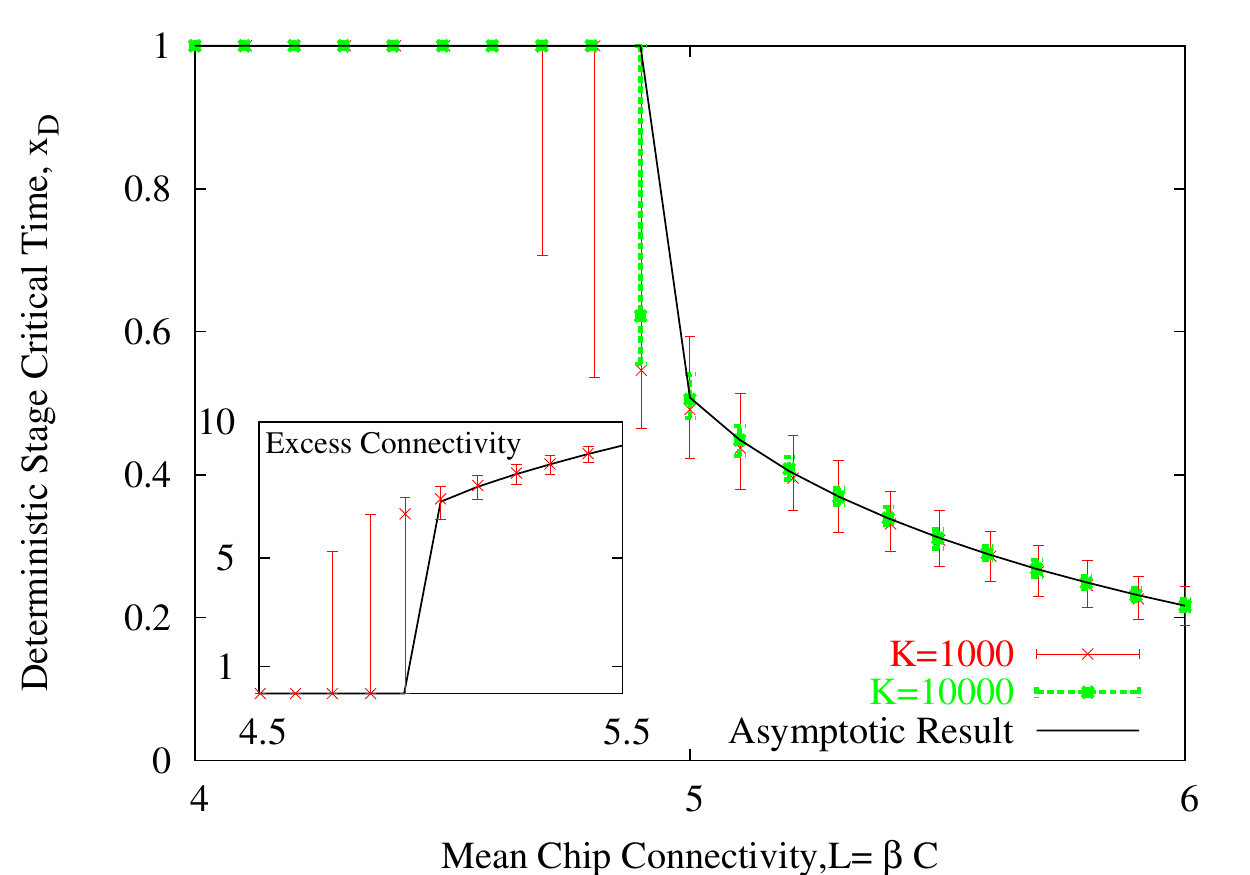}
\caption{\label{fig:ScalingC3} (colour online) The asymptotic and experimental results for $x_D$ indicates a first order transition in $\beta$ when $C\!=\!3$. The mean excess connectivity of variables remaining in the graph when $x_D<1$ is large (inset), implying the terminal residual graph is not tree like. The experimental curves indicate the median, first and third Quartiles for these values based on $1000$ ($100$) samples for for K\!=\!1000 ($10000$), typical cases are observed to concentrate on the asymptotic prediction.}
}
\end{figure}

Figure \ref{fig:ScalingC3} demonstrates the results of the numerical integration for the regular case with $C\!=\!3$ at  various loads by comparison with experiments on systems with $K\!=\!1000$ and $K\!=\!10000$. Similar excellent agreement was found for the Poissonian ensemble~\cite{Raymond:Thesis}.

\section{Computational Hardness}
\begin{figure}
\centering{
\includegraphics[width=\linewidth]{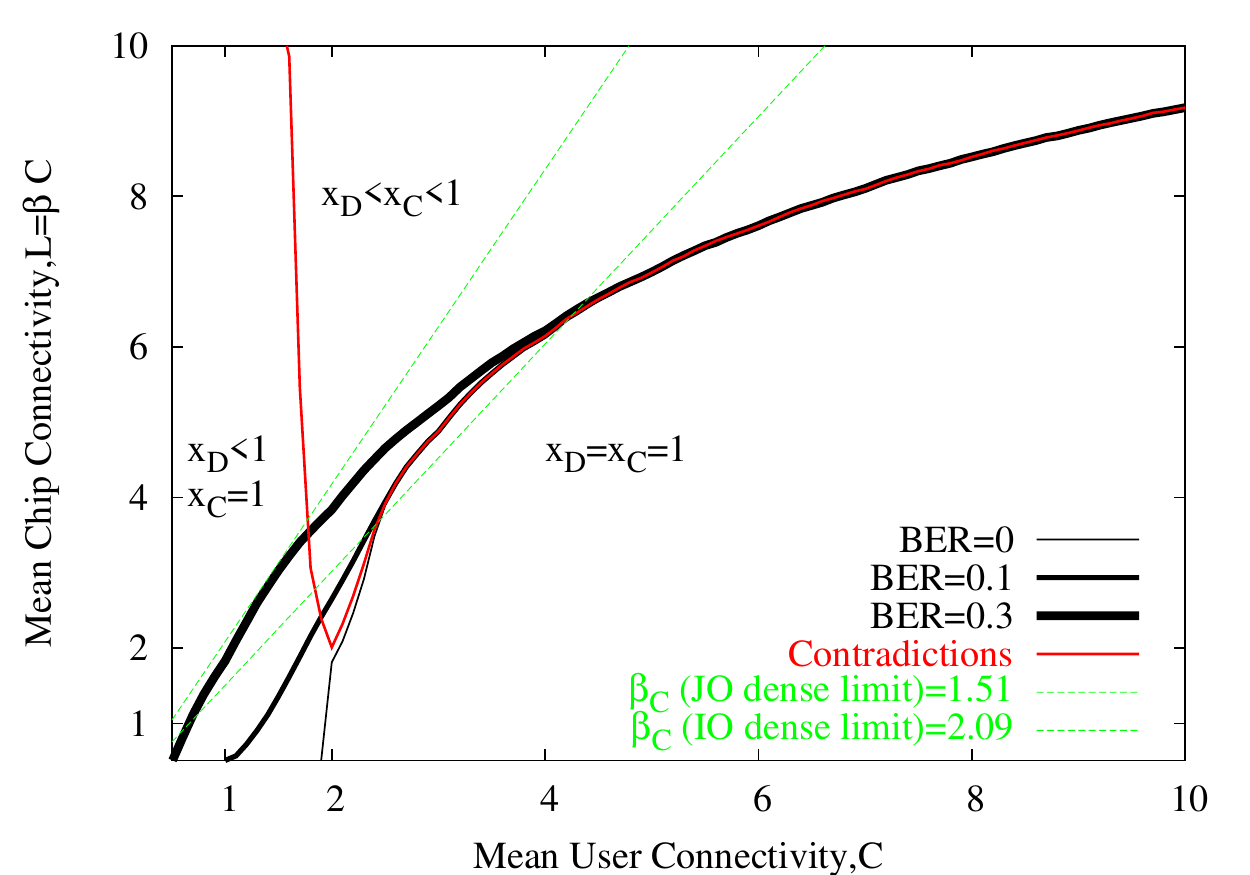}
\caption{\label{fig:Ensemble1} (colour online) Curves represent thresholds in the median of $1000$ samples for systems of $K\!=\!10000$. The regular ensemble is presented, interpolating linearly (users of connectivity $C$ and $C\!+\!1$) where $C$ is not integer. The regular ensemble produces unique solutions with $C$ greater than $2$ and $\beta$ small, and when $C\!<\!2$ (below the percolation threshold) completes without contradictions to produce one of a degenerate set of solutions without contradictions. The contours in Bit Error Rate indicate a sharp transition in estimate quality as $\beta$ increases. The limit of the regime with BER$\!=\!0$ is coincident with the line indicating presence of contradictions above the percolation threshold.}
}
\end{figure}

Figure \ref{fig:ScalingC3} shows that there is a discontinuous jump in $x_D$ as $\beta$ increases, the guessing stage goes from being empty to forming a large fraction of the algorithm run-time, and the Terminal Residual Graph (TRG) goes from empty at $x_D$ to a giant loopy graph. At small $\beta$ and $C$ greater than $2$ there is only a deterministic phase and determining the unique JO detection (also IO) is trivial ($x_D\!=\!x_C\!=\!1$). Similarly below the percolation threshold, the solution ceases to be unique,  finding one of the degenerate solutions remains easy ($x_D\!<\!1$, $x_C\!=\!1$). Finally at large $\beta$ and $C$, the algorithm involves a non-deterministic phase, and generates contradictions ($x_D\!<\!x_C\!<\!1$). This latter regime dominates for large $C$ for finite $\beta$. At large $\beta$ and finite $C$ one can expect degeneracy in the solutions, but close to the transition the embedded solution is expected to remain the unique JO solution, only difficult to find by UCP. The three regimes can be bounded in a phase diagram as shown in figure~\ref{fig:Ensemble1}. Up to the range of $\beta$ shown it is found that perfect decoding (BER$\!=\! 0$)  coincides with $x_C\!=\!1$, except at small $C$. The JO solution is unique and trivial to find at small load. \PREPRINTYES{The case of a Poissonian ensemble is examined briefly in Appendix~\ref{sec:Poissonian}.}

Figure~\ref{fig:Example63} shows properties typical of the regime in which the non-deterministic method exists.  For computational complexity reasons the gap $x_C\!-\!x_D$ is most important. If this is not asymptotically $0$, then with high probability $O(K)$ guesses have been made in reaching a contradiction. The cost of backtracking, reevaluating $O(2^K)$ combinations to find the correct branch, is not feasible and UCP becomes an efficient JO estimator.

In order to understand the algorithm failure it is useful to consider the TRG structure. In the regular case the graph consists of a C-core, every variable is connected to exactly C degenerate chips, each chip implying a constraint. These constraints are of many types, comparable to 2-XOR, 1-in-3 SAT~\cite{Raymond:PD}, and others. The clauses are {\em locked} -- at least two variables in any clause must be changed to go from one solution to another. Random Constraint Satisfaction Problems (CSP) of a similar specification have been much studied in physics~\cite{Zdeborova:LC,Raymond:PD}. Increasing the ratio of logical constraints to variables ($1/\beta$) often leads to a sharp transition in the ability to find any solution (SAT transition) buffered by a computationally hard regime. 

The case studied is topologically similar to a CSP, but the embedded solution changes the nature of the transition~\cite{Zdeborova:QP}, and it is difficult to establish the extent of this effect.  There is always a (BER$\!=\!0$) JO solution, but one may asymptotically expect a sharp transition from a computationally easy regime for JO detection, to a computationally difficult one in which the solution is still unique, and then towards regimes with many solutions (poor performance even with an ideal detector).
It may be argued that the transition is a peculiarity of the UCP decoder and the BPSK modulation scheme (which introduces degeneracy). It also seems likely that for TRGs with unique solutions methods such as BP decimation may be successful beyond the UCP threshold in typical case with some high probability.  Certainly the critical $\beta$ decreases with $C$, whereas algorithms are known to perform well up to $\beta \lesssim 1$ in practice. However, for small $C$ the thresholds for $\beta$ outlined in this paper seem to match quite well transitions in the equilibrium solution spaces for noisy systems~\cite{Raymond:SS}, which may indicate a common phenomena. \PREPRINTNO{The proposed framework can be easily extended to consider noise in the form of an erasure channel~\cite{Luby:EE}. Erasure of chips has the effect of increasing load, and inhomogeneity in the users, so that the regular code becomes qualitatively closer to the Poissonian code.}

\begin{figure}
\centering{
\includegraphics[width=\PREPRINTNO{0.85\linewidth}\PREPRINTYES{\linewidth}]{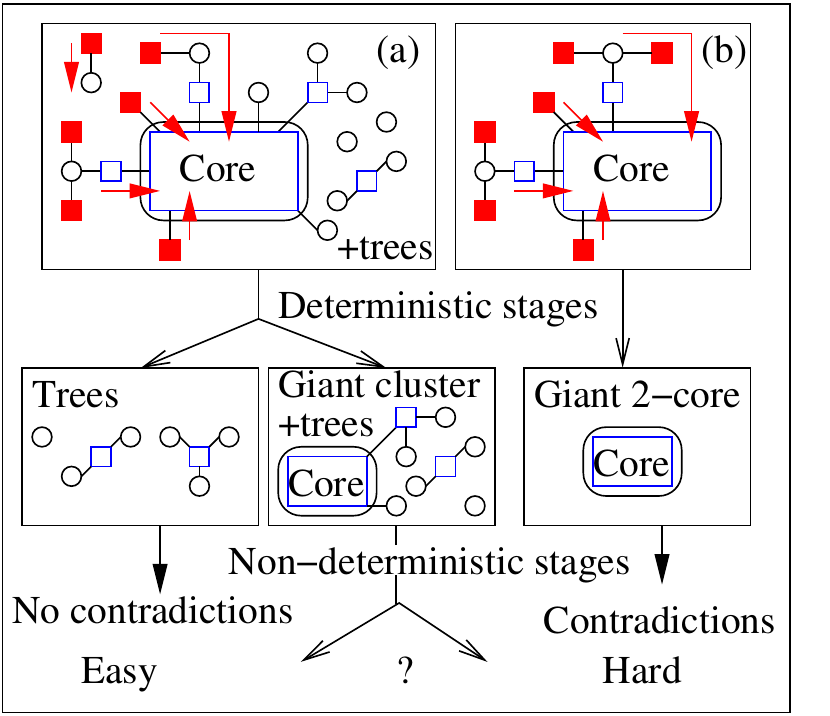}
\caption{\label{fig:figure3} (colour online) The initial problem has unit clauses and a residual graph. Information is transferred from the former to the latter by UCP. Two different inference paths correspond to the Poissonian (a) and Regular (b) problems. With erasures in the regular channel (a) becomes relevant. (a) In the Poissonian ensemble there is always an ambiguous stage due to presence of disconnected variables, but is computationally easy if only trees are present in the residual graph at $x_D$ (TRG). There are no algorithmic methods guaranteed to efficiently find solutions in worst cases where a giant component/core is present. (b) In the regular ensemble the algotihm either terminates with only a deterministic stage, or halts at a 2-core (graph without leaves). Solutions must be isolated in (b) due to the locked nature of clauses, in graphs with leaves (a) some additional freedom may apply.}
}
\end{figure}

\PREPRINTYES{
\section{The role of channel noise}

The effect of noise might be considered for an erasure channel without fundamentally changing the analysis~\cite{Luby:EE}. In an erasure channel a fraction of chips are removed, guessing becomes necessary and degeneracy is introduced. For the Poissonian ensemble the random erasure of chips is equivalent to working with a new code parameterised by larger $\beta$. However, for the sparse code any small number of erasures has an important immediate effect: the TRG 2-core cease to be the entire graph and there is no longer a unique solution (BER$\!=\!0$). The Poissonian code development scheme of figure~\ref{fig:figure3} is applicable. Nevertheless, a sharp UCP transition in $\beta$ is still observed between a computationally easy tree-like TRG (analogous to the $x_D\!=\!1$ scenario) and the non-trivial loopy TRG, except for small $C$ and/or many erasures.

Discontinuous transitions in optimal, or heuristically determined, bit error rates are common with increasing noise in CDMA detection models. Erasures within the noiseless framework introduce additional degeneracy in the solution space. It seems reasonable to expect that a critical number of erasures might cause a discontinuous jump in the BER applying to a typical jointly optimal solution, related to the solution space on the core of the TRG. An equilibrium statistical mechanics approach may be most effective at uncovering such features.
}

\PREPRINTYES{
\section{Conclusion}
Unit clauses propagation is an important component in many solution finders and seems particularly well adapted for sparse CDMA, providing a guaranteed jointly optimal performance across a range of parameters. This high performance may be established in typical case by analytic methods. In this paper it is argued that the failure of UCP in the noiseless case for large systems, due to multi-access interference at high load, characterises an asymptotically limiting regime for all fast algorithms due to fundamental topological and solution space features of the inference problem.

Understanding the residual graph problem in greater detail might provide insight for sparse codes and, if solved, a basis for selection of principled detection methods at high load. A basis for this analysis may be found in recent statistical physics research on locked constraint satisfaction problems.
}
\PREPRINTNO{
\section{Conclusion}
Results presented for noiseless sparse CDMA channels indicate new limits on the range of parameters for which optimal detection is algorithmically easy. It is argued in this paper that the inference structure undergoes a transition with increasing load, which may characterise the origins of algorithmic difficulty more generally at high load, as well as providing a limit on noisy channel optimal detection.
}


\section*{Acknowledgment}
This research is partly supported by the Research Grants Council of Hong Kong (Grant No. HKUST 604008).




%
\UPDATEBIBFILENO{


}
\UPDATEBIBFILEYES{
\bibliographystyle{IEEEtran}
\PREPRINTNO{
\bibliography{BibliographyPCN}
}
\PREPRINTYES{
\bibliography{BibliographyPCN_MAY}

\begin{thebibliography}{10}
\providecommand{\url}[1]{#1}
\csname url@samestyle\endcsname
\providecommand{\newblock}{\relax}
\providecommand{\bibinfo}[2]{#2}
\providecommand{\BIBentrySTDinterwordspacing}{\spaceskip=0pt\relax}
\providecommand{\BIBentryALTinterwordstretchfactor}{4}
\providecommand{\BIBentryALTinterwordspacing}{\spaceskip=\fontdimen2\font plus
\BIBentryALTinterwordstretchfactor\fontdimen3\font minus
  \fontdimen4\font\relax}
\providecommand{\BIBforeignlanguage}[2]{{%
\expandafter\ifx\csname l@#1\endcsname\relax
\typeout{** WARNING: IEEEtran.bst: No hyphenation pattern has been}%
\typeout{** loaded for the language `#1'. Using the pattern for}%
\typeout{** the default language instead.}%
\else
\language=\csname l@#1\endcsname
\fi
#2}}
\providecommand{\BIBdecl}{\relax}
\BIBdecl

\bibitem{Verdu:MD}
S.~Verd{\'u}, \emph{Multiuser Detection}.\hskip 1em plus 0.5em minus
  0.4em\relax New York, NY, USA: Cambridge University Press, 1998.

\bibitem{Yoshida:ASS}
M.~Yoshida and T.~Tanaka, ``Analysis of sparsely-spread \mbox{CDMA} via
  statistical mechanics,'' in \emph{Proceedings - IEEE International Symposium
  on Information Theory, 2006. (Seattle)}.\hskip 1em plus 0.5em minus
  0.4em\relax Piscataway, NJ, USA: IEEE, 2006, pp. 2378--2382.

\bibitem{Montanari:BPB}
A.~Montanari, B.~Prabhakar, and D.~Tse, ``Belief propagation based multiuser
  detection,'' in \emph{43rd Annual Allerton Conference on Communication,
  Control and Computing 2005. (Monticello)}.\hskip 1em plus 0.5em minus
  0.4em\relax Red Hook, NY, USA: Curran Associates, Inc., 2006.

\bibitem{Raymond:SS}
J.~Raymond and D.~Saad, ``Sparsely spread \mbox{CDMA} - a statistical
  mechanics-based analysis,'' \emph{J. Phys. A}, vol.~40, no.~41, pp.
  12\,315--13\,334, 2007.

\bibitem{Deroulers:CUU}
C.~Deroulers and R.~Monasson, ``Criticality and universality in the
  unit-propagation search rule,'' \emph{Eur. Phys. J. B}, vol.~49, no.~3, pp.
  339--369, 2006.

\bibitem{Rupf:OSM}
M.~Rupf and J.~Massey, ``Optimum sequence multisets for synchronous
  code-division multiple-access channels,'' \emph{IEEE Trans. on Info. Theory},
  vol.~40, no.~4, pp. 1261--1266, 1994.

\bibitem{Tanaka:SMA}
T.~Tanaka, ``A statistical-mechanics approach to large-system analysis of
  \mbox{CDMA} multiuser detectors,'' \emph{IEEE Trans. on Info. Theory},
  vol.~48, no.~11, pp. 2888--2910, Nov 2002.

\bibitem{Raymond:Thesis}
J.~Raymond, ``{Typical case behaviour of spin systems in random graph and
  composite ensembles},'' Ph.D. dissertation, Aston University, Birmingham, UK,
  November 2008.

\bibitem{Zdeborova:LC}
\BIBentryALTinterwordspacing
L.~Zdeborov\'{a} and M.~M\'{e}zard, ``Locked constraint satisfaction
  problems,'' \emph{Phys. Rev. Lett.}, vol. 101, no.~7, p. 078702, 2008.
  [Online]. Available: \url{http://link.aps.org/abstract/PRL/v101/e078702}
\BIBentrySTDinterwordspacing

\bibitem{Raymond:PD}
J.~Raymond, A.~Sportiello, and L.~Zdeborov\'{a}, ``The phase diagram of random
  1-in-3 satisfiability problem,'' \emph{Phys. Rev. E}, vol.~76, no.~1, p.
  011101, 2007.

\bibitem{Zdeborova:QP}
L.~Zdeborov\'{a} and F.~Krzakala, ``Quiet planting in the locked constraint
  satisfaction problems,'' 2009, arXiv:0902.4185.

\bibitem{Luby:EE}
M.~G. Luby, M.~Mitzenmacher, M.~A. Shokrollahi, and D.~A. Spielman, ``Efficient
  erasure correcting codes,'' \emph{IEEE Trans. on Info. Theory}, vol.~47,
  no.~2, pp. 569--584, 2001.

\end{thebibliography}
}
}
%
%

\PREPRINTYES{
\appendix
\section{The regular ensemble asymptotic analysis}
\label{sec:regularensemble}
The mean dynamics for the regular user connectivity ensemble are developed along similar lines to those presented in the main text, expanding the explanation on certain points. At decimation time $X$ typical ensemble dynamics are described sufficiently by $\Phi_l(X)$ and $\Omega_c(X)$. The latter term describes the number of variables present $c$ times in the set of unit clauses (including those that are absent $c\!=\!0$). During the deterministic phase unit clauses exist, the critical criteria demarking this algorithmic stage is
\begin{equation}
\Omega_0(X) < K-X \;.
\end{equation}
Selecting a variable independently of its degeneracy, leads to a decrease in the number of degenerate unit clauses, therefore decimation contributes to the dynamics according to
\begin{equation}
 \Delta^{(1)}\Omega_c(X)=  - \frac{\Omega_c(X)}{\sum_{c=1}^C \Omega_c(X)}\label{eq:DeltaOmega1}\;,
\end{equation}
for all non-zero $c$. The number of residual graph variables associated to a unit clause depends on its degeneracy, in the regular ensemble a variable appears $C$ times in the unit clauses and residual graph, so the average probability an instance of a variable in the residual graph corresponds to a variable decimated from the set of unit clauses, relative to a uniform sampling of the unit clauses, is
\begin{equation}
e(X) = \sum_{c=1}^C (C-c) \Omega_c(X) /\sum_{c=0}^C (C-c) \Omega_c(X) \;.\label{eq:eX}
\end{equation}
Variables decimated appear less frequently in the residual graph than the mean.

A clause population may grow with the removal of a variable, as larger degenerate clauses become smaller degenerate clauses, or become smaller as the population is reduced. A variable is coincident with a clause in proportion to its size the dynamic in expectation is
\begin{equation}
\Delta \Phi_l(X) = \frac{e(X)}{K-X}\left(- l \Phi_l(X) + (1-z_{l+1})(l+1) \Phi_{l+1}(X) \right) \label{eq:DeltaPhiReg}\;.
\end{equation}
The $\Phi_{l+1}$ dependent term is absent for the special case $l\!=\!L_{max}$. In the large system limit $L_{max}$ is not finite, but choosing a finite value generates only a very small systematic error. The factor $z_l\!=\!2^{2-l}/(1-2^{l-1})$, since it is the probability the reduced chip contains only one true or one false literal (modulated variable), $y\!=\!L-2$ or $2-L$, multiplied by the probability that the same literal is selected. Thus it is the probability a degenerate clause becomes non-degenerate when a typical variable is decimated from a clause of length $l$. The distribution of clauses of type $L,y$ is Binomial at time $X\!=\!0$, and can be shown by iteration to remain so at later times for the degenerate clauses, hence the simple form of $z_l$.

Each reduction of a large clause to a non-degenerate clause creates $l-1$ unit clauses. These new unit clauses coincide with variables in the set $\Phi_c$ in proportion to $C-c$, for example a variable $C$ times degenerate in the unit clauses cannot be represented in the graph (since it only appears $C$ times in the problem before decimation). The probability a new unit clause is coincident with a variable of degeneracy $c$ relative to the fraction of variables with degeneracy $c$ is $f_c(X)$. Each coincidence results in variables of degeneracy $c$ being transformed into variables of degeneracy $c+1$, therefore
\begin{equation}
\begin{array}{lcl}\Delta^{(2)}\Omega_c(X) \!&\!=\!&\! \left[- f_c(X) \frac{\Omega_c}{K-X} + f_{c-1}(X) \frac{\Omega_c}{K-X}\right]\\
 &\!\times\!&\! \frac{e(X)}{K-X} \sum_{l=2}^L z_l [l(l-1)] \Phi_l(X) \end{array}\label{eq:DeltaOmega2}\;,
\end{equation}
defining $f_{-1}(X)\!=\!0$ and otherwise
\begin{equation}
f_c(X) = \frac{(C-c) \Omega_c(X)}{\sum_{c=0}^C (C-c) \Omega_c(X)} \frac{\sum_{c=0}^C \Omega_c(X)}{\Omega_c(X)} \label{eq:fcX}\;.
\end{equation}
The equations (\ref{eq:DeltaOmega1}) and (\ref{eq:DeltaOmega2}) combine to produce a small change in the population $\Omega(X)$, infinitesimal by comparison with the populations up to time $x_D$. Similarly the populations of clauses of length $l$ are in expectation order $K$, but the changes due to \ref{eq:DeltaPhiReg} are $O(1)$. Taking $X\!=\!x K$, $\Phi_l(X)\!=\!K \phi_l(x)$ and $\Omega_c(X)\!=\!K\omega_c(x)$, writing $\Delta \Phi(X)$ as a Taylor expansion and keeping only leading orders in $K$ leads to a differential description of dynamics
\begin{equation}
 \frac{\rmd \phi_l(x)}{\rmd x} = \frac{e(x)}{1-x}\left(l \phi_l(x) + (1-z(l+1))(l+1)\phi_{l+1}(x) \right)\;.
\end{equation}
Similarly an equation may be written
\begin{equation}
\frac{\rmd \omega_0(x)}{\rmd x} = -\frac{f_0(x)\omega_0(x)}{(1-x)} \left[\frac{e(x)}{1-x} \sum_{l=2}^L z_l [l (l-1)] \phi_l(x)\right]\;,
\end{equation}
and for each of the variables in the unit clauses, according to their degeneracy
\begin{equation}
\begin{array}{lcl}
\frac{\rmd \omega_c(x)}{\rmd x} &=& -\frac{\omega_c(x)}{\sum_{c=1}^C \omega_c(x)} +\frac{f_{c-1}(x)\omega_{c-1}(x)-f_c(x)\omega_c(x)}{(1-x)} \\ &\times& \left[\frac{e(x)}{1-x}\sum_{l=2}^L z_l [l (l-1)] \phi_l(x)\right]\end{array}\;.
\end{equation}

As in the previous Poissonian ensemble a set of differential equations is evident, the distinction is in the necessity of the factors (\ref{eq:fcX})(\ref{eq:eX}), which can be taken as $1$ to recover the Poissonian case, and in the need to include information on the degeneracy of variables to calculate typical properties. The initial condition for $\Phi_l$ are identical to the Poissonian case (\ref{eq:Phil0}), due to the extra factor $e(x)$ these equations do not have a concise solution. Initial conditions on $\Omega_c$ (\ref{eq:Omega00}) differ from the Poissonian case, in the new model it is a Binomial distribution parameterised by $C$, and the fraction of all possible $C K$ unit clauses immediately revealed by degenerate clauses (the exponent of (\ref{eq:Omega00}) divided by $C$).

The treatment for the regular case, assuming some upper bound in the variable connectivity ($C$) and the chip connectivity ($L$) provides a basis for generalising to more complicated ensembles, although the regular and Poissonian ensembles seem most natural in the context of CDMA. The possibility to optimise ensembles based on the ability of UCP to decode is one application, similar to the application of UCP in optimisation of Linear and Fountain Codes~\cite{Luby:EE}. Solution of the equations requires numerical integration, but results with simple Runge-Kutta methods were found to be sufficient and in remarkably agreement with experiment (figure \ref{fig:ScalingC3}), verifying assumptions of the method.

\section{Asymptotic results for the Poissonian ensemble}
\label{sec:Poissonian}
\begin{figure}
\centering{
\includegraphics[width=\linewidth]{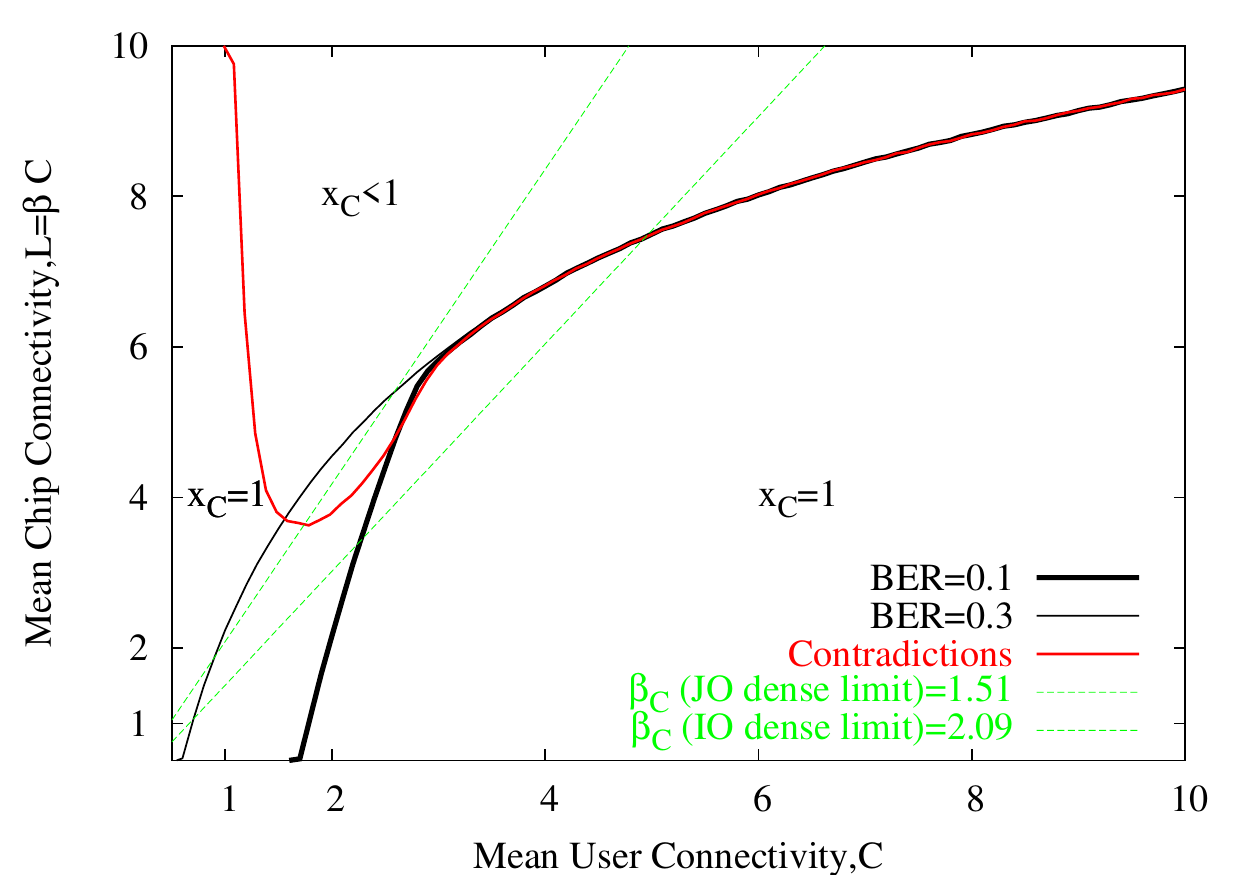}
\caption{\label{fig:Ensemble0} (colour online) A figure with statistics and parameterisations equivalent to figure \ref{fig:Ensemble1}, but applied to a Poissonian ensemble, is presented. $x_D$ is everywhere less than one, there exists three regimes distinguished by the BER of solutions and $x_C$, similar qualitatively to those of the regular code. Algorithms terminating without contradictions have low BER, except near and below the percolation transition. When $x_C$ departs from $1$ a substantial increase in BER of solutions is found.}
}
\end{figure}
A figure analogous to figure \ref{fig:Ensemble0} is shown for the Poissonian ensemble. Unlike the regular ensemble, the Poissonian ensemble nowhere results in a JO solution of BER$=0$. This is due to the inhomogeneity in user connectivity, of course only half of the unconnected users (of which an extensive number are present) may be inferred correctly for example. There is a transition between cases for which one of many JO solutions is found (without contradictions), and one in which no JO solution is found by the proposed method. A sharp transition is again seen in the structure of the TRG from a tree to a large graph containing loops at the threshold time $x_D$.

The Poissonian ensemble is not a good candidate for coded transmission. However, a regular code in combination with an erasure channel might have many properties increasingly similar to a Poissonian coded system as the noise level increases. This is observed in experiment and asymptotic analysis.

A different ensemble with regular chip-connectivity and Poissonian user-connectivity was considered in~\cite{Raymond:Thesis}. This study demonstrated that when the chip connectivity is exactly $3$ an optimal solution is always computationally easy to determine at any load. However, the transmission properties are not favourable for such a regime. This special case demonstrates the a limit with dominant dilution effect.

}

\end{document}